\documentclass[twocolumn,showpacs,preprintnumbers,amsmath,amssymb,APSl,prd,nofootinbib,superscriptaddress]{revtex4-1}

\usepackage{dcolumn}
\usepackage{bm}
\usepackage{ifpdf}
\usepackage{hyperref}
\usepackage{bm}
\usepackage{xcolor,color,graphicx,graphics}
\usepackage[spanish,english]{babel}
\usepackage[latin1]{inputenc}
\usepackage[OT1]{fontenc}
\usepackage{latexsym,amssymb,amsmath,amsfonts}
\usepackage{makeidx}
\usepackage{epsfig,subfigure}
\usepackage{natbib}
\usepackage{epstopdf}
\usepackage{mathrsfs}
\usepackage{hyperref}
\hypersetup{colorlinks=true, linkcolor=blue, citecolor=green}
\usepackage{enumerate}
\usepackage{booktabs}

\definecolor{red}{rgb}{1,0,0}

\def\+{^\dagger}

\def\<{\leftarrow}
\def\>{\rightarrow}

\def\({\left(}
\def\){\right)}

\newcommand{\bi}{\begin{itemize}} 				\newcommand{\ei}{\end{itemize}}
\newcommand{\benu}{\begin{enumerate}} 		\newcommand{\enu}{\end{enumerate}}
\newcommand{\bd}{\begin{dinglist}{0}}     \newcommand{\ed}{\end{dinglist}}
\newcommand{\bfig}{\begin{figure}[htbp]}  \newcommand{\efig}{\end{figure}}
        			
\newcommand{\bc}{\begin{center}} 				  \newcommand{\ec}{\end{center}}
\newcommand{\be}{\begin{equation}} 				\newcommand{\ee}{\end{equation}}
\newcommand{\bsub}{\begin{subequations}}  \newcommand{\esub}{\end{subequations}}
\newcommand{\ben}{\begin{eqnarray}} 			\newcommand{\een}{\end{eqnarray}}
\newcommand{\ba}[1]{\begin{array}{#1}} 		\newcommand{\ea}{\end{array}}
\newcommand{\bea}{\begin{equation}\begin{array}{rcl}}
\newcommand{\eea}{\end{array}\end{equation}}

\begin{document}

\title{Multicenter solutions in  Eddington-inspired Born-Infeld gravity}

\author{Gonzalo J. Olmo} \email{gonzalo.olmo@uv.es}
\affiliation{Departamento de F\'{i}sica Te\'{o}rica and IFIC, Centro Mixto Universidad de Valencia - CSIC.
Universidad de Valencia, Burjassot-46100, Valencia, Spain}
\affiliation{Departamento de F\'isica, Universidade Federal da
Para\'\i ba, 58051-900 Jo\~ao Pessoa, Para\'\i ba, Brazil}
\author{Emanuele Orazi} \email{orazi.emanuele@gmail.com}
\affiliation{ International Institute of Physics, Federal University of Rio Grande do Norte,
Campus Universit\'ario-Lagoa Nova, Natal-RN 59078-970, Brazil}
\affiliation{Escola de Ciencia e Tecnologia, Universidade Federal do Rio Grande do Norte, Caixa Postal 1524, Natal-RN 59078-970, Brazil}
\author{Diego Rubiera-Garcia} \email{drubiera@ucm.es}
\affiliation{Departamento de F\'isica Te\'orica and IPARCOS, Universidad Complutense de Madrid, E-28040 Madrid, Spain}

\date{\today}
\begin{abstract}
We find multicenter (Majumdar-Papapetrou type) solutions of Eddington-inspired Born-Infeld gravity coupled to electromagnetic fields governed by a Born-Infeld-like Lagrangian.  We  construct the general solution for an arbitrary number of centers in equilibrium and then discuss the properties of their one-particle configurations, including the existence of bounces and the regularity (geodesic completeness) of these spacetimes. Our method can be used to construct multicenter solutions in other theories of gravity.

\end{abstract}

\maketitle

\section{Introduction}

Among the many exact solutions of interest known within General Relativity (GR), we find the class of gravitating configurations in self-equilibrium. Such is the case, for instance, of geons \cite{Wheeler:1955zz}, gravitating solitons  \cite{Volkov:1998cc,Herdeiro:2018djx} and skyrmions \cite{Bizon:1992gb,Adam:2014dqa}, and other long-lived configurations \cite{Alcubierre:2018ahf}. There is a sub-class of solutions of such a family corresponding to those where the attractive gravitational field of a system of particles is exactly balanced by a repulsive electrostatic force. Revolving around some previous considerations by Weyl restricted to axial symmetry \cite{Weyl:1917gp}, the first explicit solution of this kind, corresponding to the sourceless Einstein-Maxwell field equations, was first found by Majumdar \cite{Majumdar} and, independently, by Papapetrou \cite{Papapetrou}. It was latter shown by Hartle and Hawking \cite{HarHaw} that the Majumdar-Papapetrou solution can actually  be interpreted as a collection of extremal black holes in static equilibrium \cite{OrtinBook}, raising further interest on the topic from different communities \cite{deBoer:2008zn,Meessen:2017rwm,Kastor:1992nn,Fernandez-Melgarejo:2013ksa}. A particularly interesting property of this solution is its lack of symmetry in the distribution of black holes. Many other solutions of this family and generalizations have been subsequently found and characterized in the literature regarding their astrophysical features \cite{Myers:1986rx,Chrusciel:1994qa,Lemos:2003gx,Frolov:2012jj,Yumoto:2012kz,Semerak:2016gfz,Bini:2019wef}.

In the quest for counterparts of gravitating configurations in  extensions of GR a number of solid scores have been hit, for instance, within rotating black holes \cite{Barack:2018yly}, compact stars \cite{Olmo:2019flu}, further horizonless compact objects \cite{Cardoso:2019rvt}, and so on. The search for such configurations is suggested from the point of view of finding alternatives to canonical GR objects (either black holes or compact stars) whose properties can be tested against the present and future stream of data from multimessenger astronomy \cite{Berti:2015itd}, and act as observational discriminators with respect to GR predictions. In implementing this program one faces the fundamental difficulty of the inherently more involved equations of motion of most extensions of GR, which largely prevents the construction of solutions of theoretical and observational interest (among other troubles). It is therefore timely and of great relevance the development of novel methods and algorithms to shortcut the structure of such field equations to find explicit solutions.

One such method has been developed recently \cite{Afonso:2018bpv}. It works for theories of gravity including scalar objects built out of contractions of the (symmetric part of the) Ricci tensor with the metric, and formulated in metric-affine (Palatini) spaces, where metric and affine connection are regarded as independent entities. The resulting family of models are the so-called Ricci-Based Gravities (RBGs), which yield second-order, ghost-free equations of motion which are compatible with all solar system and gravitational wave observations so far. For these theories it is possible to introduce new variables such that the corresponding field equations are cast in the Einstein frame, in such a way that the nonlinearities are transferred to the matter sector. From this frame one can take (when known) the corresponding GR solution, and find the counterpart in the RBG frame via purely algebraic transformations. This is known as the \emph{mapping method}, whose reliability was originally proven for anisotropic fluids \cite{Afonso:2018bpv}, and has allowed to obtain new solutions for electromagnetic \cite{Afonso:2018mxn} and scalar fields \cite{Afonso:2018hyj,Afonso:2019fzv}.

The main aim of this paper is to progress further in the analysis of the capabilities of this mapping by working out the counterpart of the Majumdar-Papapetrou (MP) solution using one particular RBG, namely, the so-called Eddington-inspired Born-Infeld (EiBI) gravity theory, popularized by Ferreira and Banados \cite{Banados:2010ix}, though first studied in the metric-affine formalism by Vollick \cite{Vollick:2003qp} inspired on a work by Deser and Gibbons \cite{Deser:1998rj}. This choice is motivated on the plethora of applications of this theory within astrophysics, black holes physics and cosmology, as found in the last few years (see  \cite{BeltranJimenez:2017doy} for a recent review). We shall explicitly cast the mapping method for this theory and the combination of fluid and electromagnetic fields supporting the MP solution, and then we will build its generalization within EiBI gravity. We shall  show that, depending on the sign of the EiBI parameter, the corresponding solutions can be understood as i) a  collection of point-like objects with  bounded stress-energy density everywhere and that lose their (extremal) horizon in the lower part of the mass spectrum, and ii) as a set of extremal black holes in equilibrium, each of which contains a non-traversable wormhole (or black bounce \cite{Simpson:2018tsi,Simpson:2019cer,Lobo:2020kxn}) in its interior. We discuss the regularity properties of both families of solutions regarding the completeness of geodesics and the behavior of curvature scalars.

The paper is organized as follows: in Sec. \ref{sec:II} we introduce the main elements of the MP solution in GR. The RBG family of theories and the mapping with electromagnetic fields is presented in Sec. \ref{sec:III}. We then combine these two ingredients to present the counterpart of the MP solution in Sec. \ref{sec:IV}, discussing its horizon structure and geodesic equation for one-particle configurations. We finally conclude in Sec. \ref{sec:V} with a discussion and some perspectives.

\section{Extreme counterpoised dust and the Majumdar-Papapetrou solution} \label{sec:II}

The Majumdar-Papapetrou family is a particular class of solutions solving Einstein equations
\begin{equation} \label{eq:EGR}
G_{\mu\nu}=\kappa^2 (T_{\mu\nu}^m + T_{\mu\nu}^{em}) \ ,
\end{equation}
where $\kappa^2=8\pi G$ is Newton's constant, while the contributions to the energy-momentum tensors read
\begin{equation} \label{eq:fluid}
T^m_{\mu\nu}=\rho\, u_{\mu}\mu_{\nu} \ ,
\end{equation}
which corresponds to a pressureless dust component, where $\rho$ is the energy density and the unit time-like vector $u^{\mu}u_{\mu}=-1$, and
\begin{equation}
T^{em}_{\mu\nu}=-\frac{1}{4\pi} \left({F^\alpha}_{\mu}F_{\alpha\nu}-\frac{1}{4}g_{\mu \nu}F^{\alpha\beta}F_{\alpha\beta} \right) \ ,
\end{equation}
which corresponds to a standard (Maxwell) energy-momentum tensor associated to the electromagnetic field, where $F_{\mu\nu}=\partial_{\mu}A_{\nu}-\partial_{\nu}A_{\mu}$ is the field strength tensor of the vector potential $A_{\mu}$. The Einstein field equations (\ref{eq:EGR})  must be supplemented with the matter field equations, which read
\begin{equation} \label{eq:Maxw}
\nabla_{\mu}F^{\mu\nu}=4\pi J^{\nu} \ ,
\end{equation}
where the charge density reads $J^{\nu}=\rho_e u^{\nu}$. This is nothing but Maxwell field equations sourced by a current generated by the pressureless fluid (\ref{eq:fluid}).

Assuming purely electric fields, $A_{\mu}=\delta_{\mu}^t \phi$, Weyl showed \cite{Weyl:1917gp} that the most general relation between the metric and the electric potential solving both the Einstein and electromagnetic equations must be of the form ($A, B$ some constants)
\be
g_{tt}=A + B\phi + \phi^2 \ ,
\ee
supported by axially symmetric spatial symmetry. The MP solution generalizes the proposal of Weyl  to any spatial symmetry by imposing the following more stringent restriction on the relation between the metric and the electrostatic potential
\be
g_{tt}=\left(C\pm \frac{\phi}{\sqrt{2}}\right)^2 \ .\label{MP-Cond}
\ee
Enforcing this condition, it can be proved that the solution is static with the density of the dust that equals the electric charge distribution, namely $\rho_e=\rho$, which is known as an extreme counterpoised dust. Moreover, the spatial sections are conformally flat when the pressure is set to zero \cite{Lemos:2005md}. Under these conditions, the MP background spacetime metric (in Cartesian coordinates) can be expressed as
\be \label{eq:MPSSS}
ds^2=-\frac{1}{U(x,y,z)^2}dt^2+U(x,y,z)^2d\vec x\cdot d\vec x \ ,
\ee
where $U$ is a function characterizing the geometry that is related to the electrostatic potential through the MP condition \eqref{MP-Cond}. Since the Einstein-Maxwell field equations only depend on derivatives of $\phi$, this potential can be redefined to cancel the constant $C$ so that \eqref{MP-Cond} simplifies as
\be
U = \pm \frac{\sqrt{2}}{\phi} \ . \label{MP-Cond-Simpl1}
\ee
This is the choice of \cite{HarHaw} up to a choice of units of the electric charge that can reabsorb the $\sqrt{2}$ factor. On the other hand, from the line element (\ref{eq:MPSSS}) the matter field equations (\ref{eq:Maxw}) result in the following Poisson equation
\be \label{eq:mateom}
\nabla^2  U=-4\pi \rho \,U^3 \ ,
\ee
that reduces to Laplace equation for the electrovacuum solution, recovering the original MP solution.

In summary, MP solutions are a particular subset  of the Einstein-Maxwell-dust system in which the mass is exactly tuned to the electric charge. For instance, any collection of extreme Reissner-Nordstr\"om black hole solutions ($m^2=q^2$) located at will is a particular MP solution \cite{OrtinBook}, without any need to impose additional symmetries. For this reason, these configurations are sometimes called {\it multicenter} solutions. In any case, of particular interest are those configurations enjoying spherical symmetry, in which case Eq.(\ref{eq:MPSSS}) can be expressed as
\be \label{eq:MP1}
ds^2=-U^{-2}(R)dt^2 +U^2 (dR^2 +R^2 (d\theta^2 + \sin^2 \theta d\phi^2)) \ ,
\ee
and thus Eq.(\ref{eq:mateom}) remains formally the same but now the Lagrangian and its functional dependencies are expressed in terms of $R$. A further coordinate change may bring the line element into standard Schwarzschild form, that is
\be \label{eq:MP2}
ds^2=-A(r)dt^2 + B(r)dr^2 + r^2 (d\theta^2 + \sin^2 \theta d\phi^2) \ ,
\ee
via the identifications $r=RU(R)$, $A(t)=U^{-2}(R)$ and $B(r)^{-1/2}=1+\frac{R}{U} dU(R)/dR$. Either under form (\ref{eq:MP1}) or (\ref{eq:MP2}), there are typically two paths followed  in the literature to solve the corresponding field equations: i) one assumes a functional form for $U(R)$ and solves (\ref{eq:mateom}) to find the matter energy density $\rho(R)$ threading the geometry \cite{Bonnor:1975gba,Kleber:2004kr} or ii) a function $\rho(R)$ for the inner region is set \cite{Horvat:2004qs}, and resort to numerical methods to resolve the corresponding differential equation in order to get $U(R)$. Perhaps the most well known example of the first path is the so-called Bonnor stars \cite{Lemos:2008pn}, where one sets two different regions as
\begin{eqnarray}
U^E&=&1+\frac{m}{r} , r \geq r_0 \\
U^I&=&1+\frac{m}{r_0} + \frac{m(r_0^2-r^2)}{2r_0 r_0^3}, 0 \leq r \leq r_0 \ ,
\end{eqnarray}
where the exterior solution, $U^E$ corresponds to an extreme Reissner-Norstr\"om black hole, which is matched to the interior solution at a certain $r=r_0$.

\section{Ricci-based gravities and the mapping for electromagnetic fields} \label{sec:III}

\subsection{Ricci-based gravities}

To describe the mapping procedure for electromagnetic fields and the main elements required for our analysis, let us begin by defining the action of RBGs as
\begin{eqnarray}\label{eq:actionRBG}
\mathcal{S}_{RBG}&=&\int d^4x \sqrt{-g}\left[ \mathcal{L}_G(g_{\mu\nu},\mathcal{R}_{(\mu\nu)}(\Gamma))+ \mathcal{L}_m(g_{\mu\nu},\psi_m)\right] \ , \nonumber
\end{eqnarray}
where $g$ is the determinant of the spacetime metric $g_{\mu\nu}$. The functional dependence of the gravitational Lagrangian $\mathcal{L}_G$ must be through traces of powers of the object ${M^\mu}_{\nu} \equiv g^{\mu\alpha} \mathcal{R}_{(\alpha\nu)}$, where $\mathcal{R}_{(\alpha\nu)}(\Gamma)$ is the symmetric part of the Ricci tensor\footnote{Note that this requirement safeguards the resulting theory from potential instabilities associated to the loss of projective symmetry induced by the antisymmetric part of the Ricci tensor, see \cite{BeltranJimenez:2019acz,Jimenez:2020dpn} for details. To lighten the notation, from now on parenthesis will be dropped.}, which is solely built out of the affine connection $\Gamma \equiv \Gamma_{\mu\nu}^{\lambda}$ (not necessarily symmetric \cite{Afonso:2017bxr}), the latter being a priori independent of the metric (Palatini approach). Regarding the matter Lagrangian $\mathcal{L}_m(g_{\mu\nu},\psi_m)$, it depends on a set of matter fields $\psi_m$ minimally coupled to the spacetime metric.

The field equations for RBGs can be conveniently written as \cite{Afonso:2018bpv,Jimenez:2020dpn}
\begin{equation}
{G^\mu}_{\nu}(q)=\kappa^2 \tilde{T}{^\mu}_{\nu}(q) \ ,
\end{equation}
where ${G^\mu}_{\nu}(q)$ is the Einstein's tensor of a new rank-two tensor related to the spacetime metric via the fundamental relation
\begin{equation} \label{eq:dm}
q_{\mu\nu}=g_{\mu\alpha}{\Omega^\alpha}_{\nu} \ .
\end{equation}
Here ${\Omega^\alpha}_{\nu}$ is dubbed as the deformation matrix, and can always be written \emph{on-shell} as a function of the stress-energy tensor $T_ {\mu\nu}(g) \equiv \frac{2}{\sqrt{-g}}\frac{\delta \mathcal{L}_m}{\delta g^{\mu\nu}}$. The relation between ${T^\mu}_\nu(g)$ and $\tilde{T}{^\mu}_{\nu}(q)$ follows from the original RBG field equations and takes the form
\begin{equation} \label{eqTmunumap}
\tilde{T}{^\mu}_{\nu}(q)= \frac{1}{|\hat{\Omega}|^{1/2}}\left[{T^\mu}_\nu(g)-\delta^{\mu}_\nu\left(\mathcal{L}_G+\tfrac{T(g)}{2}\right)\right] \ ,
\end{equation}
where vertical bars denote a determinant and $T \equiv g^{\mu}T_{\mu\nu}$ is the trace of the stress-energy tensor. This effective stress-energy tensor $\tilde{T}{^\mu}_{\nu}(q)$ can be similarly derived from another Lagrangian density $\tilde{\mathcal{L}}_m$, that is, $\tilde{T}_{\mu\nu}(q) \equiv \frac{2}{\sqrt{-q}}\frac{\delta \tilde{\mathcal{L}}_m}{\delta q^{\mu\nu}}$. This procedure establishes a correspondence or {\it mapping} between RBGs coupled to $\mathcal{L}_m$ and GR coupled to $\tilde{\mathcal{L}}_m$, as established in \cite{Afonso:2018bpv,Afonso:2018mxn,Afonso:2018hyj,Afonso:2019fzv}.  Note that, in general, both $\mathcal{L}_m$ and $\tilde{\mathcal{L}}_m$ will contain fields of the same kind (that is, scalar fields map into scalar fields, electromagnetic into electromagnetic, and so on), though the functional dependence will be different, yielding in general non-canonical Lagrangians in one (or both) sides.

\subsection{The mapping for Eddington-inspired Born-Infeld gravity}

In this work we shall be interested in the case where we choose the following RBG Lagrangian density
\begin{eqnarray}\label{GravSect}
\displaystyle{{\cal L}_{G}=\frac{1}{\epsilon\kappa^2}\left(\sqrt{\left| g_{\mu\nu}+ \epsilon \mathcal{R}_{\mu\nu}\right|} -\lambda \sqrt{-g} \right)} \ ,
\end{eqnarray}
where $\epsilon$ is a constant with dimensions of length squared, and the theory features an effective cosmological constant $\Lambda_{eff}=\frac{\lambda-1}{\epsilon}$. This is the well known and well studied Eddington-inspired  Born-Infeld (EiBI) gravity \cite{Banados:2010ix,BeltranJimenez:2017doy}. We shall first perform a few manipulations in order to apply the mapping above ({\it i.e.} Eq.(\ref{eqTmunumap})) to this setting. The equations of motion for EiBI gravity  are obtained by variation of the action (\ref{GravSect}) with respect to the connection and the metric as the two sets of nonlinear equations (here $S^\nu{}_{\alpha\gamma}$ represents the torsion tensor)
\begin{eqnarray}
\nabla_\mu\left(\sqrt{-g}\frac{\partial {\cal L}_G}{\partial \mathcal{R}_{\beta\gamma}}\right)\delta^{\mu\nu}_{\alpha\gamma} &=& \left(S^\nu{}_{\alpha\gamma}+2S^\sigma{}_{\sigma[\alpha}\delta^\nu{}_{\gamma]}\right)\sqrt{-g}\frac{\partial {\cal L}_G}{\partial \mathcal{R}_{\beta\gamma}}  \nonumber\\
2\kappa^2\frac{\partial {\cal L}_G}{\partial \mathcal{R}_{\mu\rho}}g_{\rho\nu} &=& \lambda\delta_\nu^\mu - \epsilon\kappa^2 T^\mu{}_\nu \label{MetrFieldEq} \ ,
\end{eqnarray}
respectively. Following \cite{Afonso:2017bxr} we can introduce an auxiliary metric $q_{\mu\nu}$ defined by the equation
\be\label{AuxMetr}
\sqrt{-q}q_{\mu\nu}=2\kappa^2 \sqrt{-g}\dfrac{\partial{\cal L}_G}{\partial \mathcal{R}_{\mu\nu}} \ ,
\ee
and after performing a projective transformation\footnote{For recent discussions on the interpretation of projective transformations in metric-affine gravities see \cite{Bejarano:2019zco,Jimenez:2020dpn,Garcia-Parrado:2020lpt}.}					
\be
\tilde\Gamma^\rho{}_{\mu\nu} = \Gamma^\rho{}_{\mu\nu} + \frac23 S^\lambda{}_{\lambda\mu} \delta^\rho_\nu \ ,
\ee
it is possible to reduce the field equations associated to the variation of the connection to the metric compatibility condition, $\tilde\nabla_\rho q_{\mu\nu}=0$\footnote{From now on tildes over quantities will indicate those variables defined in the GR frame; in particular this implies that indices are raised and lowered with the $q_{\mu\nu}$ metric. Conversely, when the tildes are dropped it will mean that indices are raised and lowered with the $g_{\mu\nu}$ metric instead.
}. Eq.\eqref{AuxMetr} applied to \eqref{GravSect} reveals that the auxiliary metric appearing in Eq.(\ref{eq:dm}) is
\be
\label{eq:defmat}
q_{\mu\nu}=g_{\mu\nu}+\epsilon \mathcal{R}_{\mu\nu} \ ,
\ee
which is a well known result since the seminal paper  \cite{Banados:2010ix}. Using the following general recipe of \cite{Orazi:2020mhb} to generate the metric for the EiBI gravity (\ref{GravSect}) coupled to any matter theory
\begin{equation}
g_{\mu\nu} = q_{\mu\nu} - \epsilon \kappa^2\left(\tilde T_{\mu\nu} - \frac12 \tilde T q_{\mu\nu}\right)\,,
\end{equation}
it is straightforward to find the following relation between the metrics in the case of any nonlinear electrodynamics
\begin{eqnarray}
g_{\mu\nu}&=&\left[1+\epsilon \kappa^2\left(\tilde{\cal L}_m-2\tilde{K}\frac{\partial \tilde{\cal L}_m}{\partial \tilde{K}}-\tilde{G}\frac{\partial \tilde{\cal L}_m}{\partial \tilde{G}}\right) \right]q_{\mu\nu} \nonumber \\
&+&2\epsilon \kappa^2 \frac{\partial \tilde{\cal L}_m}{\partial \tilde{K}} \tilde{K}_{\mu\nu} \ ,
\end{eqnarray}
where we have introduced the two invariants of the electromagnetic field
\be
\tilde{K}\equiv -\frac12 F_{\mu\nu}\tilde F^{\mu\nu}\,,\qquad \tilde{G}\equiv \frac14 F_{\mu\nu}{}^\star \tilde F^{\mu\nu}\,,
\ee
with ${}^\star F^{\mu\nu}=\frac{1}{2}\varepsilon^{\mu\nu\alpha\beta}F_{\alpha\beta}$  the dual of the field strength tensor $F_{\mu\nu}=\partial_{\mu}A_{\nu}-\partial_{\nu}A_{\mu}$ and $\tilde{K}_{\mu\nu} \equiv F_{\mu\alpha}{\tilde F^\alpha}_{\nu}$.  Choosing the standard Maxwell Lagrangian on the GR side, that is
\be
\label{TildeMattSect}
\tilde{\cal L}_m = \frac{\tilde K}{8\pi} \ .
\ee
the expression above boils down to
\begin{equation}
\label{eq:MetricMap}
g_{\mu\nu}=\left(1 - \frac{\epsilon \kappa^2 \tilde{K}}{8\pi}\right)q_{\mu\nu} + \frac{\epsilon \kappa^2}{4\pi}  \tilde{K}_{\mu\nu}\,.
\end{equation}
This equation provides a direct shortcut to find any solution on the EiBI side (as given by $g_{\mu\nu}$) starting from a seed solution on the GR side (as given by $q_{\mu\nu}$). In the next section we shall use this powerful result in order to generate the counterpart of the MP solution within EiBI gravity coupled to BI electrodynamics.

Let us now focus on electromagnetic fields. Following the procedure detailed in \cite{Delhom:2019zrb,Orazi:2020mhb}, the field equations \eqref{MetrFieldEq} can be reduced  to the standard Einstein equations written in terms of the auxiliary metric and the tilted connection if and only if the matter sector in the EiBI frame is related to the matter sector in the GR frame through the following parametrization \cite{Orazi:2020mhb}
\be
{\cal L}_m(g,\psi)=\frac{1}{\epsilon\kappa^2}\left\{\lambda - \frac{1-\epsilon\kappa^2\left(\tilde{\cal L}_m(g,\psi)-\tilde T/2\right)}{\sqrt{\det{\left[\delta^\mu_\nu - \epsilon\kappa^2\left(\tilde{\cal L}_m(g,\psi)-\tilde T/2\right)\right]}}}\right\}
\ee
that reduces to (cfr. Eq.(5.11) of Ref.\cite{Delhom:2019zrb}) (here $\tilde{\kappa}^2\equiv \kappa^2/8\pi$)
\be \label{eq:lmhat}
{\cal L}_m = \frac{1}{\epsilon \kappa^2}\left[\frac{(\epsilon \tilde{\kappa}^2 \tilde K-1)}{1-(\epsilon \tilde{\kappa}^2)^2\left(\tilde K^2 + 4\tilde G^2\right)} + \lambda\right] \ ,
\ee
provided that we choose the matter content on the GR side as given by Maxwell electrodynamics in \eqref{TildeMattSect}.
Our next goal is to express \eqref{eq:lmhat} in terms of quantities in the EiBI frame by writing the invariants of the $q_{\mu\nu}$ frame, which are those appearing in the Lagrangian density (\ref{eq:lmhat}), in terms of those of the RBG frame (the untilted variables). Using the inverse mapping between metrics (\ref{eq:defmat}), a {\it little} algebra allows to find the following relations between the field invariants in the GR and RBG frames \cite{Orazi:2020mhb}:
\begin{eqnarray}
\label{Param}
\tilde K&=&\frac{2\left( K+4\epsilon\tilde{\kappa}^2 G^2\right)}{\epsilon^2\tilde{\kappa}^4\left(K^2 + 4 G^2\right)}
\nonumber \\
&\times&\left(1-\epsilon\tilde{\kappa}^2 K \pm \sqrt{1-2\epsilon\tilde{\kappa}^2 K-4\epsilon^2\tilde{\kappa}^4 G^2}\right) \\
\tilde G &=& -\frac{2G}{\epsilon^2\tilde{\kappa}^4\left(K^2 + 4 G^2\right)}\sqrt{1-2\epsilon\tilde{\kappa}^2 (K + 2\epsilon\tilde{\kappa}^2 G^2)}
\nonumber\\
&\times&\left(\sqrt{1-2\epsilon\tilde{\kappa}^2 (K + 2\epsilon\tilde{\kappa}^2 G^2)}\pm (1-\epsilon\tilde{\kappa}^2 K) \right) \ .
\end{eqnarray}
Replacing these expressions into the parametrization (\ref{eq:lmhat}) one gets the form of the matter Lagrangian in the EiBI frame as
\be
{\cal L}_m = \frac{\left(2\lambda -1\right) \pm \sqrt{1-2\epsilon\tilde{\kappa}^2 (K+ 2\epsilon\tilde{\kappa}^2 G^2)}}{\epsilon\kappa^2} \ .
\ee
Now, if we take from now on, for simplicity, asymptotically flat solutions, $\lambda=1$, and make the identification $\beta^2 = 4\pi/(\epsilon\kappa^2)$ choosing the minus sign solution, then the Lagrangian density above becomes
\begin{eqnarray}\label{MattSect}
{\cal L}_{m}=\frac{\beta^2}{4\pi}\left(1-\sqrt{1-\frac{K}{\beta^2} - \frac{G^2}{\beta^4}}\right) \ ,
\end{eqnarray}
which is the well known Born-Infeld (BI) theory of electrodynamics. Therefore, we have just seen that GR coupled to Maxwell electrodynamics maps into EiBI gravity coupled to BI electrodynamics. Note, however, that the sign of $\epsilon$ is not {\it a priori} restricted to be positive, which implies that $\beta^2$ could also be negative despite being written as the square of a parameter $\beta$.

\section{Mapping multicenter solutions} \label{sec:IV}

\subsection{The Majumdar-Papapetrou solution in EiBI}

Now that we have all the necessary elements of the mapping for this scenario under control, the derivation is remarkably simple. Indeed, in order to extract the counterpart of the MP solution  (\ref{eq:MPSSS}) within EiBI gravity  we just need to compute the extra corrections appearing in Eq.(\ref{eq:MetricMap}) via the ansatz (\ref{eq:MP1}). First we find that
\be\label{eq:eigenvectors}
\tilde{K}_{\mu \nu}dx^\mu dx^\nu=
\frac{(\nabla U)^2}{U^6} dt^2 -\frac{U_idx^iU_j dx^j}{U^2} \ ,
\end{equation}
where $\{i,j\}=x,y,z$, which allows to find
\be\label{eq:eigenvectors}
\tilde{K} =-\frac{(\nabla U)^2}{U^4} \ ,
\end{equation}
so that after a bit of algebra Eq.(\ref{eq:MetricMap}) reads
\begin{eqnarray}
ds^2=&-&\frac{1}{U^2}\left(1+\frac{\epsilon \kappa^2}{16\pi} \frac{(\nabla U)^2}{U^4}\right) dt^2 +\frac{\epsilon \kappa^2}{8\pi} \frac{(dU)^2}{U^2} \nonumber  \\
&+&  U^2 \left(1-\frac{\epsilon \kappa^2}{16\pi} \frac{(\nabla U)^2}{U^4}\right) d\vec{x}^2 \ , \label{eq:dscart}
\end{eqnarray}
which is the counterpart of the MP solution, given by a certain $U(\vec{x})$, within EiBI gravity (\ref{GravSect}) coupled to Born-Infeld electrodynamics (\ref{MattSect}), as obtained via the mapping. Nice and easy.

To construct multicenter solutions out of the above solution one can consider a collection of $N$ solutions of this type such that the metric function (in GR) is written as
\begin{equation} \label{eq:multiGR}
U=1+\sum_i^N \frac{m_i}{\sqrt{(\vec{x}-\vec{x}_i)^2}}
\end{equation}
where $m_i$ labels the mass of each object, and $(\vec{x}-\vec{x}_i)^2=(x-x_i)^2+(y-y_i)^2+(z-z_i)^2$, where $(x_i,y_i,z_i)$ are the coordinates labeling the center of each solution. If instead of Cartesian coordinates one labels the centers by means of spherical coordinates, each vector $\vec{x}_i$ becomes $\vec{x}_i=r_i(\sin\theta\cos\varphi,\sin\theta\sin\varphi,\cos\theta)$ such that
\begin{equation}
(\vec{x}-\vec{x}_i)^2=r^2+r_i^2-2 r r_i (\sin\theta\sin\theta_i\cos(\theta-\theta_i)+\cos\theta\cos\theta_i)
\end{equation}
Inserting back the ansatz (\ref{eq:multiGR}) into the line element (\ref{eq:dscart}) one would find the counterpart of the multicenter solutions of GR. However, the resulting such expressions are not very illuminating, so in the next section we shall further constrain this setting.

\subsection{Features of the solutions near individual center}

For the sake of the discussion of the features of the generalized MP solutions, it is much more convenient to rewrite the generalized MP solution (\ref{eq:dscart}) in terms of the electrostatic potential. This can be done after noting that the expression of the electrostatic potential $A_{\mu}(x,y,z)=(\phi(x,y,z),\vec{0})$ with $\phi(x,y,z)=1/U(x,y,z)$  allows to write $\vec{\nabla}U/U^2=-\vec{\nabla}\phi$, which simplifies many expressions. Replacing this into the line element (\ref{eq:dscart}), and suitably rearranging terms one finds
\begin{eqnarray} \label{eq:lineiso}
ds^2&=&-\phi^2 \Big(1+\frac{\epsilon \kappa^2}{16\pi}  (\nabla{\phi})^2\Big) dt^2+\frac{\epsilon \kappa^2}{8\pi\phi^2}  (\vec{\nabla}\phi \cdot d\vec{r})^2  \\
&+&\frac{1}{\phi^2} \Big[\Big(1-\frac{\epsilon \kappa^2}{16\pi}  (\nabla{\phi})^2\Big)(dr^2+r^2d\Omega^2) \Big] \nonumber \ ,
\end{eqnarray}
where $d\Omega^2=d\theta^2+\sin^2 \theta d\varphi^2$ is the usual unit volume element of the two-spheres. In order to understand the properties of this multicenter solution, it is useful to have a careful look at individual centers, which will provide a reasonable approximation of the geometry for sufficiently separated objects. Corrections depending on the separation could be computed by perturbative methods and the general case should take into account the exact line element (\ref{eq:lineiso}).

In the case of having only one center, the angular contributions to the anisotropy vanish and only the radial part remains, allowing us to write $\vec{\nabla}\phi \cdot d \vec{r}=\phi_r dr$. The line element (\ref{eq:lineiso}) can then be written as
\begin{eqnarray} \label{eq:ds2}
ds^2&=& \left[1+\frac{\epsilon \kappa^2}{16\pi} \phi_r^2 \right]\left(-\phi^2  dt^2 +\frac{1}{\phi^2}  dr^2\right) \nonumber \\
&+& \left[1-\frac{\epsilon \kappa^2}{16\pi} \phi_r^2 \right]\frac{r^2}{\phi^2}d\Omega^2 \ .
\end{eqnarray}
Given that for one-particle configurations one has the electrostatic potential $\phi=r/(r+m)$, where $m$ is its mass, it is easy to go from the current isotropic coordinates to Schwarzschild-like ones by redefining the factor $r^2/\phi^2=R^2=(r+m)^2$, which turns the above one-particle line element into
\begin{eqnarray}\label{eq:1center}
ds^2&=& \left[1+s \frac{R_c^4}{R^4} \right]\left(-\phi^2(R)  dt^2 +\frac{1}{\phi^2(R)}  dR^2\right) \nonumber \\
&+& \left[1-s\frac{R_c^4}{R^4}\right]R^2d\Omega^2 \ ,
\end{eqnarray}
where now $\phi(R)=1-m/R$, the parameter $s=\pm 1$ denotes the sign of $\epsilon$, and $R_c^4\equiv\frac{|\epsilon| \kappa^2m^2}{16\pi}$ becomes a fundamental scale that characterizes the features of  these solutions.

A glance at the factor that multiplies the two-spheres in the above line element shows that if $s=+1$ then the area of the two-spheres vanishes at $R=R_c$, which suggests to introduce a new coordinate $\rho^2=R^2-R_c^4/R^2$ that allows to rewrite the line element in such a way that the center would be located at $\rho=0$. By contrast, for $s=-1$ we find that the two-spheres have a non-vanishing minimum area at $R=R_c$, which can be interpreted as representing the throat of a wormhole which satisfies the flare-out condition for sustainability (see e.g. \cite{Kim:2013tsa},  while for a detailed account of wormhole physics see \cite{VisserBook}) or, alternatively, as a black bounce  \cite{Simpson:2018tsi,Simpson:2019cer,Lobo:2020kxn}. The two signs, therefore, describe quite different objects, namely, point-like particles if $s=+1$ and wormholes if $s=-1$. In light of this observation, in the following sections we will study the properties of these one-center extremal solutions\footnote{For an exhaustive discussion of the one-center solutions away from extremality and with rotation, see \cite{rotating}.} considering the two cases $s=\pm1$ separately.

\subsubsection{Curvatures of individual centers} \label{sec:curvatures}

A look at the curvature scalars provides a useful comparison with the GR solutions. In GR ($s=0$), the Ricci scalar vanishes and the Kretschmann diverges at $R=0$ as $\sim m^4/R^8$.
When $s=+1$, we find that as $R\to R_c$ the Ricci scalar goes like $\sim 1/(R-R_c)^2$ and the Kretschmann as $\sim 1/(R-R_c)^4$, having a softer behavior if $R_c=m$, where they become $ (2m)^{-1}/(R-R_c)$ and $ (2m)^{-2}/(R-R_c)^2$, respectively.
If $s=-1$, we find that as $R\to R_c$ the Ricci scalar goes like $\sim 1/(R-R_c)^3$ and the Kretschmann as $\sim 1/(R-R_c)^6$, having also a softer behavior if $R_c=m$, where they become $-\frac{3}{2} (2m)^{-1}/(R-R_c)$ and $ \frac{9}{4}(2m)^{-2}/(R-R_c)^2$, respectively. Since this $s=-1$ case represents a wormhole, with its throat at $R=R_c$, it is also relevant to look at the curvature scalars in the limit $R\to 0$, where the area of the two-spheres goes to infinity again. In this region, we find that both the Ricci and the Kretschmann scalars are finite, taking the values $36m^2/R_c^4$ and $408m^4/R_c^8$, respectively.  We thus see that in all cases the relevant curvature divergences of these solutions are much weaker than in the GR case.

\begin{table*}
   \begin{tabular}{| c | c | c | c | c | c | c | c |}
        \hline
     \footnotesize   s & \footnotesize Type &\footnotesize Ricci Scalar &\footnotesize Kretschmann &\footnotesize Horizons &\footnotesize Surf. Grav. $\kappa$ &\footnotesize $F_{\mu\nu}F^{\mu\nu}$ &\footnotesize $\rho$ \\ \hline
     & & & & & & & \\
 \footnotesize $0$ & \footnotesize Point-like &\footnotesize $0$ &\footnotesize $\sim \frac{1}{R^8}$  &\footnotesize Double at $R=m$ &\footnotesize  $0$ &\footnotesize $\frac{m^2}{R^4}\to \infty$ &\footnotesize  $\frac{m^2}{R^4}\to \infty$  \\
  & & & & & & & \\ \hline
  & & & & & & & \\
 \footnotesize  & &\footnotesize  $\sim \frac{1}{(R-R_c)^2}$ if $R_c>m$  &\footnotesize   $\sim \frac{1}{(R-R_c)^4}$ if $R_c>m$
  & \footnotesize Double at $R=m$  &  & &  \\
  $+1$ &\footnotesize Point-like &  & & \footnotesize if $m>R_c$. & \footnotesize $0$ & \footnotesize Finite at $R=R_c$  & \footnotesize $\infty$ at $R=R_c$  \\
  & & \footnotesize $\sim \frac{1}{(R-R_c)}$ if $R_c=m$& \footnotesize $\sim \frac{1}{(R-R_c)^2}$ if $R_c=m$   & \footnotesize Naked otherwise.   & & & \\
  & & & & & & & \\ \hline

    & & & & & & & \\
  & &\footnotesize $\sim \frac{1}{(R-R_c)^3}$ if $R_c>m$  &\footnotesize  $\sim \frac{1}{(R-R_c)^6}$ if $R_c>m$
  & \footnotesize Double at $R=m\neq R_c$  & \footnotesize  $\infty$ at $R_c\neq m$ & &  \\
   \footnotesize $-1$ &\footnotesize Wormhole &  & && & \footnotesize  $\infty$ at $R=R_c$ & \footnotesize Finite at $R=R_c$   \\
  & & \footnotesize $\sim \frac{1}{(R-R_c)}$ if $R_c=m$& \footnotesize $\sim \frac{1}{(R-R_c)^2}$ if $R_c=m$   & \footnotesize Single at $R=R_c$   & \footnotesize $0$ otherwise & & \\
  & & & & & & & \\ \hline

     \end{tabular}
   \caption{Summary of the features of the one-center solutions of GR ($s=0$) and of the two families of configurations studied in this paper ($s=\pm1$). The case in which $R_c=m$ is considered separately in Sec.\ref{Rc=m}.}
 \label{table:I}
\end{table*}

\subsubsection{Horizons of individual centers}

For static individual centers, spherical symmetry allows us to identify the location of horizons by finding $R=$constant hypersurfaces with vanishing norm.
Regarding such horizons, the case $s=+1$ has the same structure as the GR solution, with a degenerate horizon located at $R=m$, whenever $m>R_c$ (equivalently $m>(|\epsilon|\kappa^2/16\pi)^{1/2}$). If $m<R_c$ (equivalently $m<(|\epsilon|\kappa^2/16\pi)^{1/2}$)  then there is no degenerate horizon because the area of the two-spheres vanishes at $R=R_c>m$ and the geometry cannot be extended further below.  Note that for such small mass configurations if one assumes that $|\epsilon|\sim l_{Planck}^2$, then $m\lesssim m_{Planck}$. The case $R_c=m$ presents a degenerate horizon at $R=m$ of vanishing area.

\begin{figure}[t!]
\centering
\includegraphics[width=0.45\textwidth]{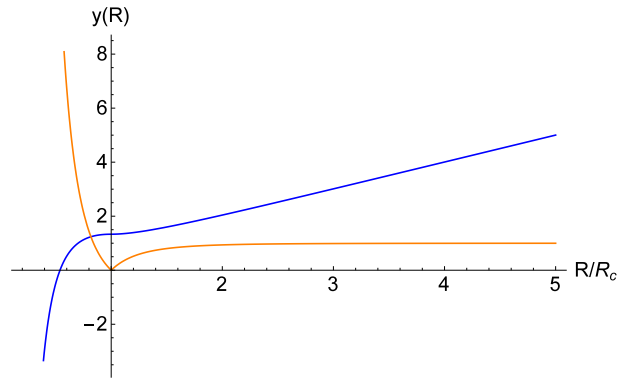}
\caption{The change of coordinates (\ref{eq:coc}) for $y(R)$ (blue) and its derivative $y'(R)$ (orange). Note that both functions are everywhere continuous. \label{fig:cv}}
\end{figure}

In the case $s=-1$,   it is important to note that precisely at $R=R_c$, where the two-spheres reach their minimum, the $g_{tt}$ component vanishes while $g^{RR}$ diverges, which requires a more careful analysis. The situation can be clarified if we transform the line element (\ref{eq:1center}) to ingoing Eddington-Finkelstein coordinates via the change $dv\equiv dt+f(R)dR$, with $f(R)=\pm 1/\phi(R)$, such that the line element (\ref{eq:1center}) becomes
 \begin{eqnarray}\label{eq:1center-EFx}
ds^2=& -&\left[1+s\frac{R_c^4}{R^4} \right]\phi^2  dv^2 -2\left[1+s\frac{R_c^4}{R^4}  \right]dv  dR \nonumber \\
&+& \left[1-s\frac{R_c^4}{R^4}\right]R^2d\Omega^2 \ .
\end{eqnarray}
Now, in order to avoid the coordinate singularity at $R=R_c$ due to the vanishing of the off-diagonal term $dv dR$ when $s=-1$, a further change of variables is necessary to absorb the factor $\left[1-\frac{R_c^4}{R^4}  \right]$ in a new radial coordinate. For this purpose we consider the choice
\begin{equation} \label{eq:ycoord}
dy=\pm\left[1-\frac{R_c^4}{R^4}  \right]dR \ ,
\end{equation}
with the plus sign corresponding to $R>R_c$ and the minus sign to $R<R_c$ to guarantee that $y(R)$ is a monotonic function on all the domain $R\in ]0,\infty[$ (see Fig.\ref{fig:cv}). The explicit relation between $y$ and $R$ can be integrated as
 \begin{equation} \label{eq:coc}
 y=\left\{\begin{array}{lr} R+\frac{R_c^4}{3R^3} & \text{ if } R\ge R_c \\
 \frac{8}{3}R_c-\left(R+\frac{R_c^4}{3R^3}\right) & \text{ if } 0<R\le R_c \end{array}\right. \ ,
 \end{equation}
with $y(R_c)\equiv y_c=4R_c/3$. This construction guarantees the continuity and derivability of the change of coordinates, as can be seen in Fig.\ref{fig:cv}. Note that in the neighborhood of $R_c$, one has $y\approx y_c\pm \frac{2}{R_c}(R-R_c)^2$.  As a result of this change of coordinates, (\ref{eq:1center-EFx}) becomes
  \begin{eqnarray}\label{eq:1center-EFy}
ds^2=& -&\left[1-\frac{R_c^4}{R^4(y)} \right]\phi^2  dv^2 -2dv  dy \nonumber \\
&+&\left[1+\frac{R_c^4}{R^4(y)}\right]R^2(y)d\Omega^2 \ ,
\end{eqnarray}
which covers the whole range $y\in ]-\infty,+\infty[$ with a single chart and avoids coordinate  metric singularities everywhere (except, obviously, in the angular sector).

 In these coordinates one finds that the normal vector to the hypersurfaces $y=$constant becomes null at $R=m$ (if $m\ge R_c$) and at $R=R_c$ (always). The time-like Killing vector $\chi=\partial_v$ also has vanishing norm there, confirming that these two locations represent Killing horizons (if $m\ge R_c$).

\subsubsection{Surface Gravity}

The properties of the horizons in the $s=-1$ model require further attention because while at $R=m$ one generically finds zero surface gravity, this quantity diverges at $R_c$ as
\begin{equation}
|\kappa| \approx \underset{R\to R_c}{\lim} \frac{({R_c}-m)^2}{2{R^2_c} (R-{R_c})} \ .
\end{equation}
The reason lies on the fact that the surface gravity is defined as the limit $|\kappa| = \underset{y\to y_H}{\lim}|A_y|$, where $A(y)= \left[1-\frac{R_c^4}{R^4(y)} \right]\phi^2$ and $y_H$ represents the location of the horizon. Since near the $R=R_c\neq m$ horizon one finds that $A(y)\approx 4(R-R_c)/R_c=\sqrt{(y-y_c)/2R_c}$ has a square root dependence on $(y-y_c)$, then its derivative necessarily induces a divergence in the denominator of $\kappa$. Only when $R_c=m$ does $\kappa$ vanish at $R_c$ because then $A(y)\sim (y-y_c)^{3/2}$. The divergence of the surface gravity in this one-center solution is a generic property that only disappears when a specific charge-to-mass relation is satisfied, which includes the case $R_c=m$ but also other cases with non-vanishing temperature  \cite{rotating}.

 To deepen into the physical meaning of this infinite surface gravity, it might be useful to have a look at the properties of the matter field there. Given that for the GR solution one has $\tilde{K}=m^2/R^4$, one readily finds that
 \begin{eqnarray}
 K&=& \frac{m^2}{R^4}\frac{1}{\left(1+s\frac{R_c^4}{R^4}\right)^2} \\
 \mathcal{L}_m&=& \frac{m^2}{16\pi R_c^4}\frac{(R_c^4/R^4)}{\left(1+s\frac{R_c^4}{R^4}\right)} \\
 \rho&=& \frac{m^2}{16\pi R_c^4}\frac{(R_c^4/R^4)}{\left(1-s\frac{R_c^4}{R^4}\right)} \ ,
 \end{eqnarray}
 where $\rho$ represents the field energy density.

 When $s=-1$, it is evident that both the electric field intensity squared and the electromagnetic Lagrangian (\ref{eq:lmhat}) diverge at $R=R_c$, but the energy density is always finite and positive.

 On the other hand, for $s=+1$ all those quantities have the reversed behavior, namely, the electric field intensity squared and electromagnetic Lagrangian (which is related to the transversal pressures of the field) are finite and well behaved, though the energy density diverges. We thus see that the divergence of curvature scalars at $x=R_c$ is totally uncorrelated with the behavior of the matter fields, which may or may not be divergent at that location.  It is thus unclear if there is any physical reason or implication for the divergence of the surface gravity at the wormhole throat when $R_c\neq m$.

The features discussed above of both $s=\pm 1$ cases regarding the behaviour of curvature scalars, horizons, surface gravity, electric fields, and energy density, have been summarized in Table \ref{table:I}, together with their comparison with GR expectations.

\subsubsection{Geodesic structure of individual centers}\label{sec:geodesics}

For spherically symmetric spacetimes  the geodesic equation can be written in a simple form \cite{Olmo:2016tra}, which for the line element (\ref{eq:1center}) becomes
 \begin{equation} \label{eq:geoeq}
 \left(\left[1+s\frac{R_c^4}{R^4}\right]\frac{dR}{du}\right)^2=E^2-V_{eff}(R(y)) \ ,
 \end{equation}
where $u$ is here the affine parameter and $E$ the energy per unit mass. As usual, this geodesic equation is akin to the motion of a single particle in a one-dimensional effective potential, which in the present case reads explicitly
 \begin{equation} \label{eq:Veff}
 V_{eff} = \Bigg(1+s\frac{R_c^4}{R^4}\Bigg)\Bigg(1-\frac{m}{R}\Bigg)^2 \left(\frac{L^2}{R^2-s\frac{R_c^4}{R^2}}-k\right) \ ,
 \end{equation}
with $k=0,-1$ for null and timelike geodesics, respectively, and $L$ denotes the angular momentum per unit mass.

\begin{figure}[t!]
\includegraphics[width=0.47\textwidth]{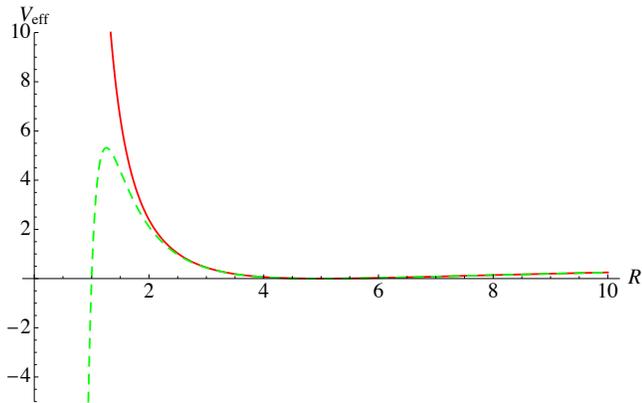}
\caption{The effective potential $V_{eff}$ in Eq.(\ref{eq:Veff}) for timelike ($k=-1$) radial ($L = 0$) geodesics for $m/R_c=5$ as a function of $R/R_c$. The solid (red) curve represents the case $s=+1$ while the dashed (green) curve is for $s=-1$. Note that in this figure the wormhole throat is located at $R=R_c=1$, where $V_{eff}^-$ vanishes.  \label{fig:Veff}}
\end{figure}

Let us consider first the case $s=+1$. As we approach the center, $R\to R_c$, the effective potential is dominated by the term
\begin{equation}
V_{eff}\approx \frac{L^2 (R_c-m)^2}{2R_c^3(R-R_c)} \ ,
\end{equation}
which represents a divergent barrier for any nonzero angular momentum [see Fig.\ref{fig:Veff}] as long as $R_c\neq m$. Thus, light rays and massive particles will bounce before reaching the center and will never get there. For exactly radial motions, $L=0$, then $V_{eff}\approx -2k(R_c-m)^2/R_c^2$, which vanishes for null rays and becomes a finite positive barrier for time-like observers ($k=-1$). Thus, particles with $E^2>2(R_c-m)^2/R_c^2$ will be able to reach the center in finite affine time.

If $m>R_c$ it is unclear if the curvature divergence at the center will do any harm to infalling observers because physical objects have a finite extension and, as we have just seen, any nonzero $L$ will experience a bounce before reaching the center. One expects that the internal forces that keep the object cohesive will slow down the infinitesimal elements falling radially ($L=0$) and make them bounce with the rest of the body. Nonetheless, a detailed analysis of the behavior of geodesic congruences and/or the interaction of waves with these objects would be necessary to further explore the regularity of these solutions \cite{Olmo:2016fuc}. In any case, the fact that the energy density is divergent at the center suggests a physical obstruction to the extension of null radial geodesics in these configurations.

Let us now consider the wormhole case, $s=-1$.
The shape of the effective potential is depicted in Fig.\ref{fig:Veff} for the case of time-like ($k=-1$) radial ($L=0$) geodesics, though we point out that for geodesics with non-zero angular momentum the qualitative behaviour is similar. Only those geodesics with energy $E$ larger than the maximum of the effective potential will be able to go to the interior region of these solutions and interact with the wormhole throat. In the limit $m\gg R_c$ this maximum is located at $R\approx 3^{1/4}R_c$ and grows as $V_{max}\approx \frac{2 m^2}{3 \sqrt{3}R_c^2}$. In general, at $R=R_c$ we have $V_{eff}=0$ [see Fig.\ref{fig:Veff}], leading there to ($u_c$ being an integration constant)
\begin{equation}\label{eq:geonearWH}
\pm E(u-u_c) \approx  (y-y_c) \ ,
\end{equation}
for all null (with angular momentum) and time-like geodesics (recall Eq.(\ref{eq:ycoord})). This solution is exactly the same as the one corresponding to null radial geodesics ($k=0, L=0$), for which $V_{eff}=0$ everywhere, and indicates that null rays and massive particles reach the surface $R=R_c$ in finite affine time. Whether geodesics can be extended beyond this point is a subtle issue with no straightforward answer. On the one hand, Eq.(\ref{eq:geonearWH}) shows that the relation between the radial coordinate $y$ and the affine parameter $u$ is smooth across the throat. However, one could argue that the divergence of curvature scalars (and of the surface gravity) at $R_c$ should preclude the extensibility of geodesics, though there are examples which contradict this view \cite{Olmo:2016fuc,Olmo:2015bya,Bazeia:2015uia,Olmo:2015dba} (see also \cite{Bejarano:2017fgz,Menchon:2017qed,Bambi:2015zch}) based on the fact that the energy density at $R_c$ is finite. In addition, it has been shown in \cite{rotating} that radial null geodesics in the one-center solutions are insensitive to the charge and mass parameters, existing a specific combination of them for which all curvature scalars are finite. Since there are no reasons to believe that geodesics should not be extended in that specific case and the geodesics satisfy exactly the same equation, it seems natural to conclude that they are always extensible across $x=R_c$.

\begin{figure}[t!]
\includegraphics[width=0.45\textwidth]{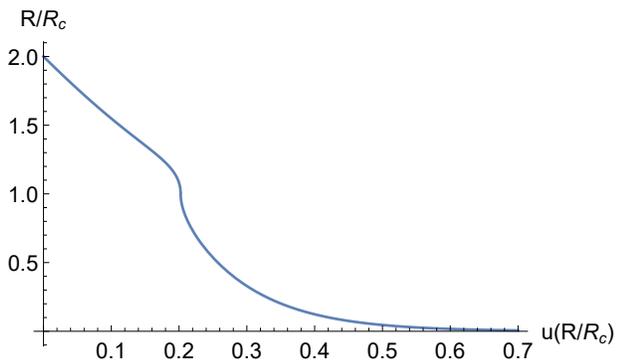}
\caption{The affine parameter for timelike ($k=-1$) radial ($L = 0$) geodesics as a function of $R/R_c$ (note the inversion of axis), obtained by integration of the geodesic equation (\ref{eq:geoeq}) with the effective potential (\ref{eq:Veff}), taking $E=6$, $m/R_c=10$ and $u_c=2$. Note that in this figure the wormhole throat is located at $R=R_c=1$. As is apparent from this figure, $u$ can be indefinitely extended across the wormhole throat towards $u \rightarrow + \infty$ at $R=0$. \label{fig:affnul}}
\end{figure}

\begin{table*}
   \begin{tabular}{| c | c | c | c | c | c | c | c |}
        \hline  & & & &  \\
     \footnotesize   s &\footnotesize Time-like $L\neq 0$ &  \footnotesize Time-like $L=0$ &  \footnotesize Null $L\neq 0$ & \footnotesize Null $L=0$  \\

     & & & &  \\ \hline
  0    & \footnotesize Complete & \footnotesize Complete & \footnotesize Complete &  \footnotesize Incomplete \\
       & & & &  \\ \hline
        & & & &  \\
       & \footnotesize Complete if $E^2<\frac{2(R_c-m)^2}{R_c^2}$& & &  \\
  +1    & & \footnotesize Complete &\footnotesize Complete &\footnotesize Incomplete \\
        &\footnotesize Incomplete otherwise & & &  \\
        & & & &  \\ \hline
        & & & &  \\
  -1   &\footnotesize Complete$^*$ &\footnotesize Complete$^*$ & \footnotesize Complete$^*$& Complete \\
  & & & & \\ \hline
     \end{tabular}
   \caption{Properties of the geodesics of the one-center solutions of GR ($s=0$) and of the two families of configurations studied in this paper ($s=\pm1$). For clarifications on what Complete$^*$ means, see the discussion in Sec.\ref{sec:geodesics}. }
 \label{table:II}
\end{table*}

In Fig.\ref{fig:affnul} we illustrate the case of time-like ($k=-1$) radial ($L=0$) geodesics assuming that they can be extended across the wormhole throat $R=R_c$ (note that there is no numerical problem in doing so).

Provided that we have continued the geodesics across $R=R_c$, to see what happens with them  in the asymptotic region $R\to 0$ we need to study the effective potential (\ref{eq:Veff}) that for massive particles goes like
\begin{equation}
V_{eff}\approx -\frac{m^2 R_c^4}{R^6} \ ,
\end{equation}
while for null non-radial geodesics  ($k=0, L\neq 0$) one has
\begin{equation}
V_{eff}\approx -\frac{L^2m^2 }{R^4} \ .
\end{equation}
In both cases, the effective potential is attractive and divergent.
In the latter case, one finds that the geodesic equation  (\ref{eq:geoeq}) can be integrated as
\begin{equation}\label{eq:nullgeo}
u=\mp\frac{R_c^3}{m L R} \ ,
\end{equation}
while for the former ($k=-1$) we have instead
 \begin{equation}\label{eq:timegeo}
u=\pm\frac{R_c^2}{m}\ln R \ ,
\end{equation}
where the minus sign in (\ref{eq:nullgeo}) corresponds to outgoing geodesics and to ingoing geodesics in (\ref{eq:timegeo}). Both of the above expressions show that as $R\to 0$  the affine parameter diverges, $u \to \pm \infty$ [see Fig.\ref{fig:affnul}], confirming in this way that all such geodesics are complete from that side since they would take an infinite affine time to get there.

Let us finally point out that the case of radial null geodesics is trivial, since from Eq.(\ref{eq:geoeq}) one immediately finds that
\begin{equation}
y=\pm Eu\ ,
\end{equation}
and $y\in ]-\infty,\infty[$ (as follows from (\ref{eq:coc})). Thus, the region corresponding to $R\to 0$ (or $y\to -\infty$) represents a regular boundary of this spacetime, since it cannot be reached in finite affine time by any geodesic.

Table \ref{table:II} summarizes the above discussion on the geodesic completeness of both $s=\pm 1$ cases, together with their comparison with GR expectations.

\subsubsection{The limiting case $R_c=m$.}\label{Rc=m}

\begin{figure}[]
\centering
\includegraphics[width=0.25\textwidth]{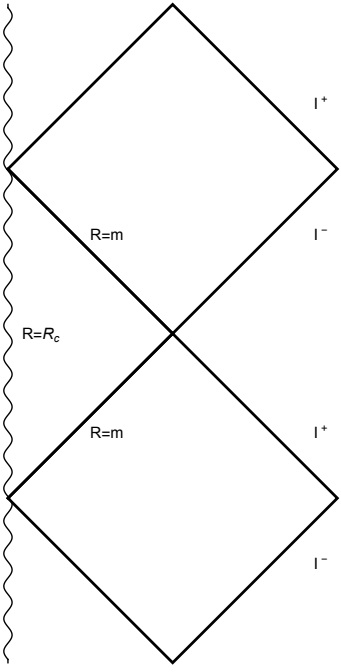}
\caption{Conformal diagram of the case $m>R_c$ when $s=+1$. The diagram repeats itself in the vertical axis indefinitely. The wavy line represents the location of curvature divergences.\label{fig:PD_p1_2sqrs}}
\end{figure}

\begin{figure}[]
\centering
\includegraphics[width=0.25\textwidth]{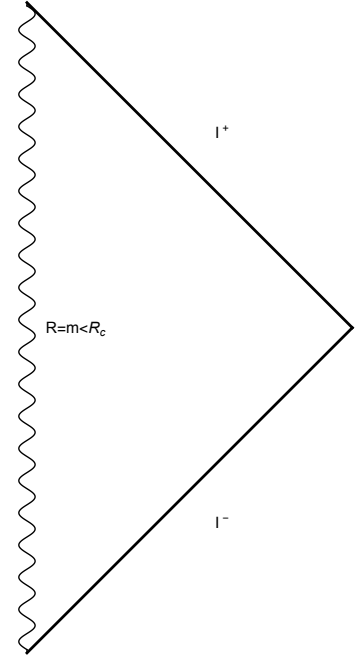}
\caption{Conformal diagram of the case $m<R_c$ when $s=+1$. The center of the object becomes a naked singularity. \label{fig:PD_p1_triangle}}
\end{figure}

\begin{figure}[]
\centering
\includegraphics[width=0.45\textwidth]{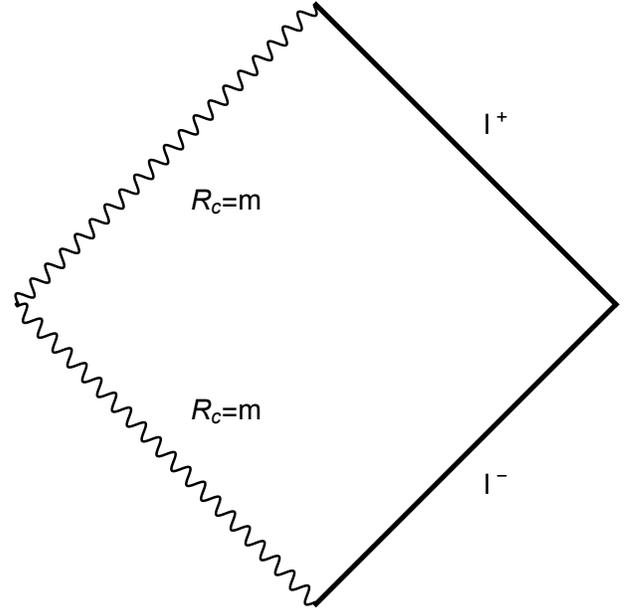}
\caption{Conformal diagram of the case $m=R_c$ when $s=+1$. The center of the object becomes a null hypersurface. \label{fig:PD_p1_RceqM}}
\end{figure}

This case deserves particular attention because it represents a limiting situation with peculiar features. From (\ref{eq:1center-EFx}) it is easy to see that for $s=+1$ the center of the object is located at $R=R_c=m$ and represents a degenerate horizon. This fact has a deep impact on the corresponding Penrose diagram of this configuration. In fact, when $m>R_c$ the basic building block of the space-time global structure can be depicted by the diagram in Fig. \ref{fig:PD_p1_2sqrs}, where curvature divergences arise at $R=R_c$ (time-like wavy line), while the two degenerate horizons are found at $R=m$. If $m>R_c$ then the curvature divergences at $R=R_c$ become naked, with no degenerate horizons, as shown in Fig. \ref{fig:PD_p1_triangle}, while for $R_c=m$ we find an intermediate situation, with the center $R=R_c$ becoming a degenerate horizon (see Fig.\ref{fig:PD_p1_RceqM}). Regarding geodesics, the case $R_c=m$ is quite different from the others with $R_c\neq m$. In the latter, the potential barrier diverges as $R\to R_c$, forcing all $L\neq 0$ geodesics to bounce before reaching it. However, when $R_c=m$, the potential barrier goes to zero linearly as $R-R_c$ as the center is approached. Thus, all geodesics with $L\neq 0$ and enough energy to overcome the maximum of the centrifugal barrier will be able to get to $R=R_c$ and hit the curvature divergence. Note, in this respect, that the strength of the divergence in this case is much weaker than in GR (see Table \ref{table:I}).

\begin{figure}[]
\centering
\includegraphics[width=0.45\textwidth]{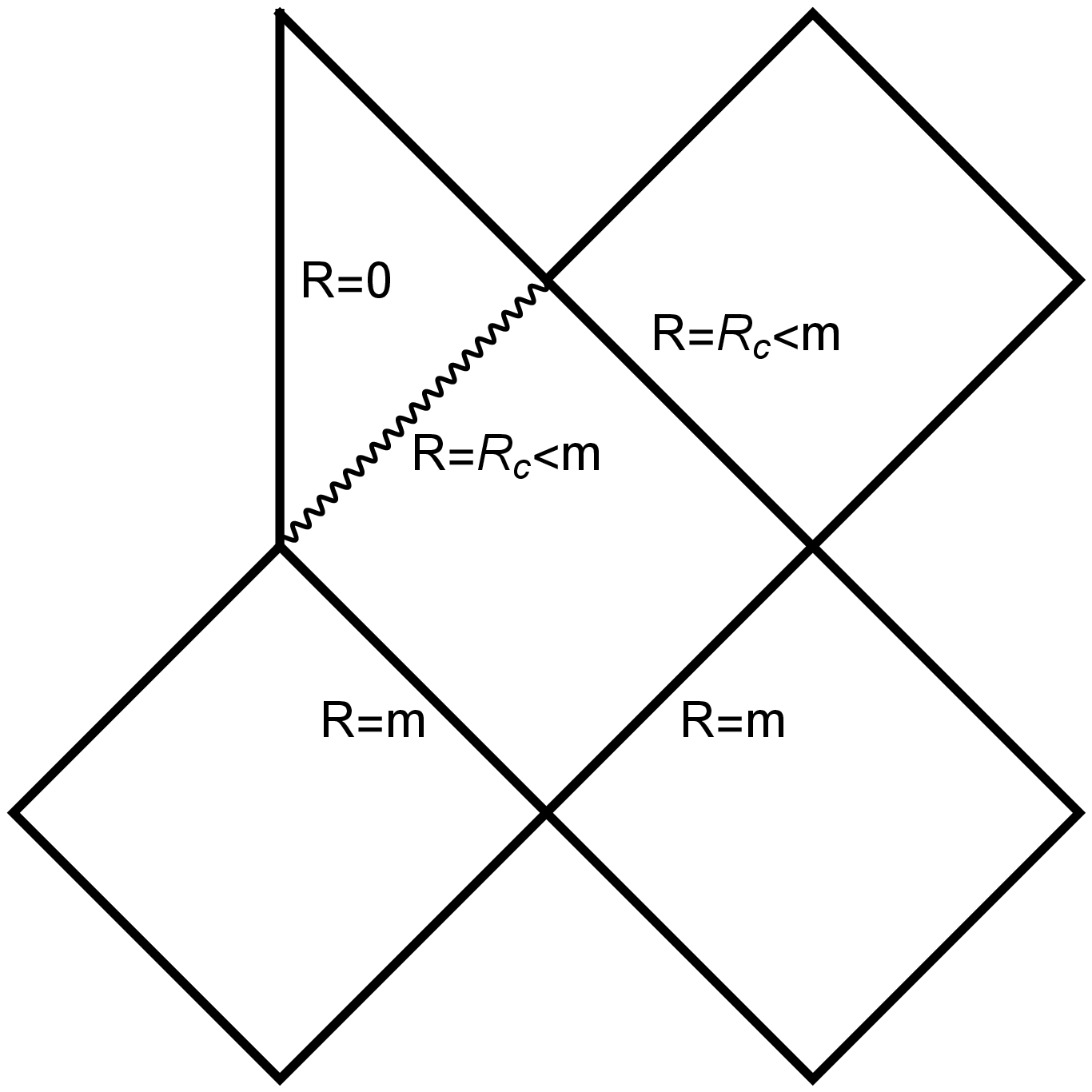}
\caption{Conformal diagram of the case $m>R_c$ when $s=-1$. The diagram repeats itself in the vertical axis indefinitely. The wavy line represents the location of curvature divergences. \label{fig:PD_m1_4sqrs}}
\end{figure}

\begin{figure}[]
\centering
\includegraphics[width=0.35\textwidth]{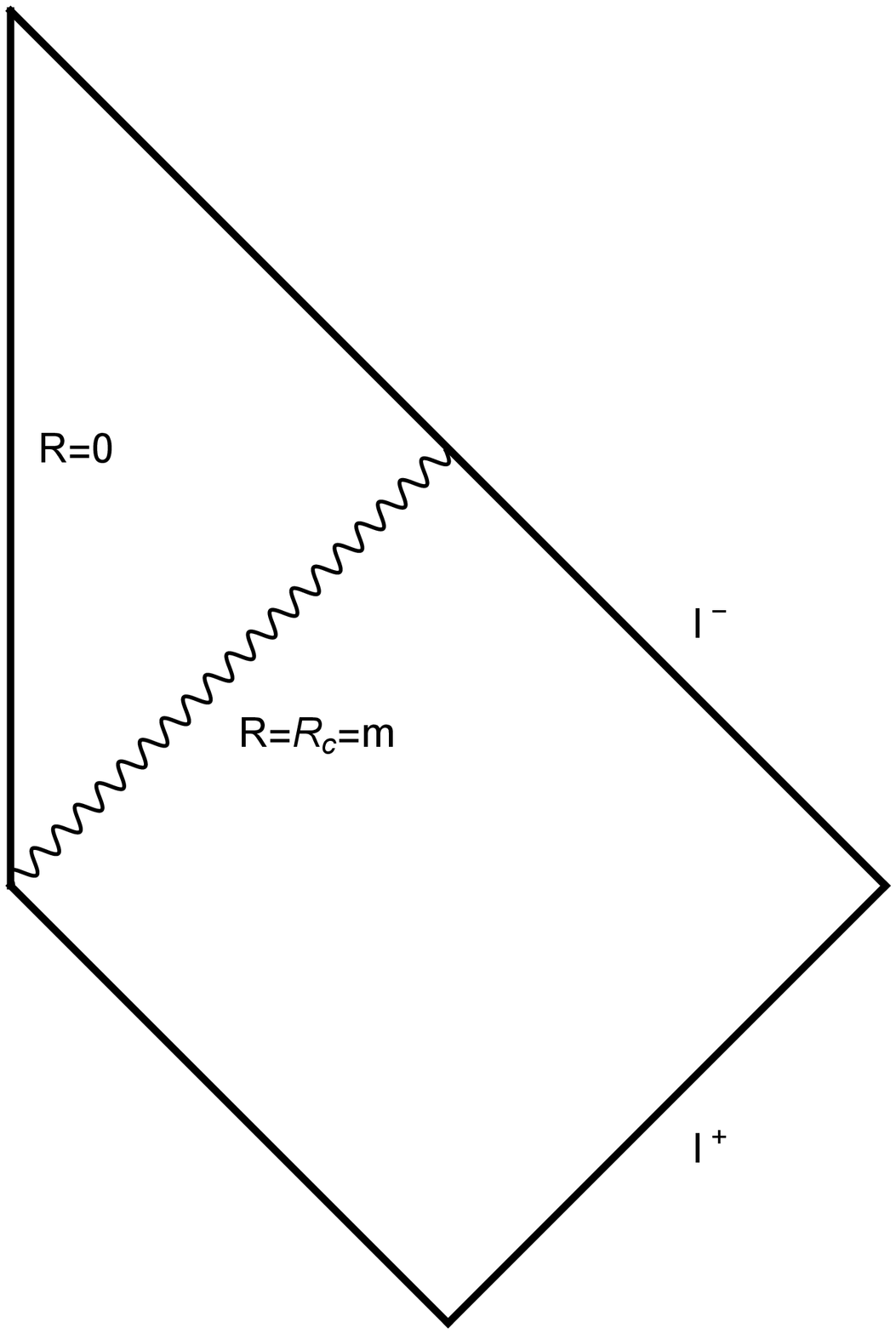}
\caption{Conformal diagram of the case $m=R_c$ when $s=-1$. The diagram repeats itself in the vertical axis indefinitely. The wavy line represents the location of curvature divergences. \label{fig:PD_m1_1sqr}}
\end{figure}

When $s=-1$ the global structure of the solutions is completely different. As we already saw, if $R_c\neq m$ both $R=m$ and $R=R_c$ are Killing horizons, with the degenerate horizon $R=m$ happening before $R=R_c$ if $m>R_c$ and after otherwise, as shown in Fig.\ref{fig:PD_m1_4sqrs}, which illustrates only the case $m>R_c$. The critical case $R_c=m$ leads to a completely different scenario, shown in Fig. \ref{fig:PD_m1_1sqr}, in which the horizons shrink to a line with a curvature divergence at the wormhole throat, which is now located at $R=R_c=m$. The behavior of geodesics in this case is qualitatively the same as discussed above for the general case.

\section{Conclusions} \label{sec:V}

In this work we have considered a particular class of the Majumdar-Papapetrou family of solutions of GR, given by a collection of extreme black holes in equilibrium (multicenter solutions), to build a new family of such objects within modified gravity. The gravitational model considered here is the Eddington-inspired Born-Infeld gravity, a particular member of the Ricci-based gravities class, while to generate such solutions we have made use of a new powerful tool dubbed as the mapping method. We have shown that via this method, solutions of the Einstein-Maxwell system are mapped into those of the EiBI gravity theory coupled to a Born-Infeld-type electrodynamics. We then used the MP multicenter solution on the GR side as the seed to generate a new family of solutions in our modified gravity model. The resulting solutions represent a collection of exotic new objects in equilibrium with important novelties as compared to those of GR, whose properties critically depend on the sign of the EiBI parameter $\epsilon$.

On the one hand, when $\epsilon>0$, one finds a family of point-like objects which look like extremal black holes for masses above the Planck scale but which loose their horizon  in the lower part of the mass spectrum. These objects have bounded electric field at their centers but divergent energy density.

On the other hand, the interpretation of the case  $\epsilon<0$ requires to play a bit to cast the corresponding line element in  suitable coordinates. There we have shown that for the case of individual centers they are a kind of extremal black hole with an internal wormhole, whose throat represents a Killing horizon with divergent surface gravity and curvature. In the interior region, geodesics take an infinite affine time to get to $R\to 0$, which represents a boundary of infinite area. We have interpreted these solutions as geodesically complete because of results previously obtained in \cite{rotating}, though the coexistence of a finite energy density at the throat with a divergent electric field intensity is certainly inconvenient. Let us point out that asymmetric wormhole solutions with incomplete geodesics have been found in the EiBI theory coupled to scalar fields \cite{Afonso:2019fzv,Afonso:2017aci}, but in such a case the matter fields at the wormhole throat are well behaved while it is the innermost region which becomes problematic. These results defy the intuitive notion that wormholes always allow for the completeness of geodesics, and raises further questions on the suitability of different markers to characterize pathologies in the characterization of spacetimes (for a broad discussion on singularity regularization see \cite{Carballo-Rubio:2019fnb,Carballo-Rubio:2019nel}). The question on the stability of the solutions presented here  is intimately related to whether the supersymmetry of the GR solution is preserved or broken under the mapping transformation \cite{Gibbons:1982fy,Tod:1983pm}. If the solutions of the EiBI theory remain supersymmetric and represent the corresponding Bogomol'nyi bound, stability would be guaranteed \cite{Brooks:1995wb}. This is an aspect that lies beyond the scope of the current work and will be explored elsewhere.

Using the methods presented in this work, it is possible to generate multicenter solutions in other gravity theories, such as $f(\mathcal{R})$ and others, coupled to nonlinear electrodynamics theories whose form depends on the specific target gravity theory chosen. One thus may wonder if it would be possible to find multicenter solutions in new gravity theories coupled to Maxwell electrodynamics. This question is relevant because, in particular, in the EiBI case coupled to Maxwell electrodynamics, individual wormhole solutions are known and are better behaved than those found here \cite{Olmo:2016fuc,Olmo:2015bya,Bazeia:2015uia,Olmo:2015dba} (some of them are traversable, not hidden behind an event horizon). Thus there is the hope of generating multiwormhole traversable configurations on which ideas related to quantum entanglement and the ER=EPR correlation could be tested \cite{Lobo:2014fma}. Attempts to find such solutions have also been carried out by the authors but without success so far. The key reason for this negative result seems to lie on the lack of specific symmetries in the field equations of the Einstein+nonlinear electrodynamics system. On the other hand, applications of these new multicenter configurations within the context of supersymmetric theories is yet to be explored.  Research in these directions is ongoing and we hope to be able to report on them elsewhere.

\section*{Acknowledgements}

DRG is funded by the \emph{Atracci\'on de Talento Investigador} programme of the Comunidad de Madrid (Spain) No. 2018-T1/TIC-10431, and acknowledges further support from the Ministerio de Ciencia, Innovaci\'on y Universidades (Spain) project No.  PID2019-108485GB-I00/AEI/10.13039/501100011033, and by the Funda\c{c}\~ao para a Ci\^encia e
a Tecnologia (FCT, Portugal) research projects Nos. PTDC/FIS-OUT/29048/2017 and PTDC/FIS-PAR/31938/2017.   This work is supported by the Spanish project  FIS2017-84440-C2-1-P (MINECO/FEDER, EU), the project H2020-MSCA-RISE-2017 Grant FunFiCO-777740, the project PROMETEO/2020/079 (Generalitat Valenciana), and the Edital 006/2018 PRONEX (FAPESQ-PB/CNPQ, Brazil, Grant 0015/2019). This article is based upon work from COST Actions CA15117 (\emph{Cosmology and Astrophysics Network for Theoretical Advances and Training Actions}) and CA18108 (\emph{Quantum gravity phenomenology in the multi-messenger approach}), supported by COST (European Cooperation in Science and Technology).

\end{document}